\documentclass[doublecol]{epl2}
\usepackage{times}
\usepackage{graphicx}
\usepackage{amsmath}
\usepackage{amssymb}

\graphicspath{{./}} 
\DeclareGraphicsExtensions{.eps}

\title{Dynamics of forced biopolymer translocation}

\author{V.V. Lehtola \and R.P. Linna \and K. Kaski}
\shortauthor{V.V. Lehtola \etal}

\institute{
  Department of Biomedical Engineering and Computational Science,\\
  Helsinki University of Technology,\\
  P.O. Box 9203, FIN-02015 TKK, Finland
  }


\abstract{
We present results from our simulations of biopolymer translocation in
a solvent which explain the main experimental findings. The
forced translocation can be described by simple force balance
arguments for the relevant range of pore potentials in experiments and
biological systems. Scaling of translocation time with polymer length
varies with pore force and friction. Hydrodynamics affects this
scaling and significantly reduces translocation times.
}

\begin{document}

\maketitle

The transport of biopolymers through a nano-scale pore in a membrane
is a ubiquitous process in biology. For example, in protein import
into mitochondria, chloroplasts, and peroxisomes the translocation
occurs with the aid of a membrane
potential~\cite{alberts}. Experimental work on forced (or
biased) translocation is largely motivated by finding methods for
reading the DNA and RNA sequences. These nanopores are
typically either fabricated solid-state~\cite{storm,li} or
$\alpha$-hemolysin ($\alpha-$HL) pores in lipid bi-layer
membranes~\cite{meller,kasianowicz}. Foundation for the theoretical
work was laid in the classic treatment by Sung and Park~\cite{sung},
which was based on the assumption that the polymer segments on the two
sides of the membrane reside close to separate thermal
equilibria. However, the validity of this approach was questioned
already in~\cite{chuang,kantor}, where the authors noted that the
pore force regime in which the polymer's relaxation
time towards equilibrium is smaller than the characteristic
translocation time is marginal and that the approach would
  be  invalid even in the unforced translocation for sufficiently long
polymers. Theoretical work, inconsistent with experiments, has since
evolved in different directions. 

The role of computer simulations has been largely to support the theoretical
work which neglects hydrodynamics. Hence, results from simulations where
hydrodynamic interactions are included are few and, due to their being
computationally demanding, often fairly qualitative~\cite{ali,gauthier}. In
addition, the generally used Monte Carlo method gives unphysical behaviour
for larger pore force values relevant for experiments and biological
systems~\cite{storm,meller,kasianowicz}, as we have
shown~\cite{ourpre}. Very recently multi-scale simulations on
biopolymer translocation in a solvent were reported to give results in
accordance with experiments~\cite{bernaschi, fyta_pre}.

Our motivation for the present study is two-fold. First, by using realistic
dynamics we want to find explanation for the dynamics of the
experimentally observed translocation processes. Secondly, we want to
determine the effect of hydrodynamics on forced polymer translocation,
previously studied only in the {\it unforced}
case~\cite{ali,gauthier}. We use a hybrid multi-scale method, where
the polymer
follows detailed molecular dynamics and the coarse-grained solvent
stochastic rotation dynamics (SRD). The solvent is divided
into cells, within which fictitious solvent particles perform
simplified dynamics where collisions among them and with the polymer beads
are taken effectively into account by performing random rotations of
the random part of their velocities, $v_i(t+\Delta t_{SRD}) = \mathbf{R}
[v_i(t) - v_{cm}(t)] + v_{cm}(t)$, where $v_i$ are the
particle velocities  inside a cell, $\Delta t_{SRD}$ is the time step
for solvent dynamics, $\mathbf{R}$ is the rotation matrix, and $v_{cm}$
is the centre-of-mass velocity of the particles within the
cell. Hydrodynamic modes are supported over the cells. Optionally,
they can be switched off by not adding back $v_{cm}$ after the random
rotation, which is particularly feasible for pinning down the effect
of hydrodynamics. The above-described collision step is followed by
the free-streaming step $r_i(t + \Delta t_{SRD}) = r_i(t) + v_i(t)
\Delta t_{SRD}$. Thermostating is done
by rescaling all solvent particle velocities so that
equipartition theorem is fulfilled at all times. More detailed
descriptions of the method can be found {\it e.g.}
in~\cite{malevanets,ihlekroll,webster}.

In this paper we study the forced
translocation where the two sides separated by walls are not
hydrodynamically coupled. To achieve this we use a non-aqueous pore,
{\it i.e.} there are no solvent particles inside the pore. This
corresponds closely to the experiments we aim to model and also
addresses the theoretical predictions, where the two subspaces separated
by the wall are taken to be uncoupled. In addition, technically
speaking the
coarse-grained solvent dynamics does not allow for overly confined
spaces, but implementing solvent dynamics in dimensions smaller than
the SRD cell dimension would require detailed molecular dynamics for
the solvent. The linear SRD cell dimension in our model is $\Delta x
\equiv 1.0 \equiv b$, where $b$ is the polymer bond length.

The standard bead-spring chain is used as a coarse-grained
polymer model~\cite{ourpre,linna}. Adjacent
monomers are connected with anharmonic springs, described by the
finitely extensible nonlinear elastic (FENE) potential,
\begin{equation}
U_{FENE} = - \frac{K}{2} R^2 \ln \big ( 1- \frac{r^2}{R^2} \big ).
\label{fene_mc}
\end{equation}
Here $r$ is the length of an effective bond, $R = 1.5$ the maximum bond
length. The Lennard-Jones (LJ) potential 
\begin{eqnarray}
U_{LJ} &=& 4 \epsilon \left[ \left(\frac{\sigma}{r}\right)^{12} -
\left(\frac{\sigma}{r}\right)^{6} \right]
, \: r \leq 2^{-1/6} \sigma \nonumber \\
U_{LJ} &=& 0 , \; r > 2^{-1/6} \sigma,
\label{lj_md}
\end{eqnarray}
is used between all beads. The parameter values were chosen to be $\epsilon
= 1.2$, $\sigma = 1.0$, and $K = 60 / \sigma^2$. The used LJ potential with
no attractive part mimics good solvent condition for the polymer.

We do not include a harmonic bending potential, which would change the
above-described freely-jointed chain (FJC) to the worm-like-chain
(WLC). Elastically, the FJC and WLC were seen to differ only
marginally and their hydrodynamic characteristics were found
identical in the present model~\cite{linna}. The swelling exponents
measured for the FJC and WLC in the present model were identical,
which means that
since packaging of polymers is not an issue in the present geometry,
as {\it e.g.} in capsids~\cite{ali_packaging}, the bending potential
does not constitute an important factor to the translocation process.

The model geometry contains a slit\footnote{ The slit mimics
experimental setups including confining walls. The slit is large enough not
to affect dynamics of the translocating polymer.} formed by two walls
perpendicular to the $x$-direction. Periodic boundary conditions are applied
in $y$- and $z$-directions. A third wall, three polymer segments, $b$, thick
and impermeable to the solvent, is placed in the middle of the system,
perpendicular to the $z$ direction.  No slip boundary conditions are applied
between the three walls and the solvent. A circular nanopore of diameter
$1.2\ b$ is placed in the centre of the middle wall. The force $f$ acting on
the beads inside the nanopore is constant and local for the pore, which
models well the experimental setups and biological systems, where solvents
are good ionic conductors eliminating any potential gradients outside the
pore. The polymer beads inside the pore are not coupled with
hydrodynamic modes or with the heat bath and in the directions perpendicular
to the cylindrical pore walls experience a damped harmonic potential $U_h$,
described by $- \nabla U_h = F_h = -k r_{x,y} - c v_{x,y}$, where $k=1000$,
$c=10$, $r_{x,y}$ is the polymer bead position with respect to the centre
axis of the cylindrical pore, and $v_{x,y}$ is the velocity component
perpendicular to the pore walls. Thus $U_h$ centres the polymer along the
$z$-directional axis of the pore. The potential is chosen large, so no
hairpin configuration can enter the pore as its width is effectively small.
Hence, the polymer segment inside the pore remains rather straight. In the
$z$ direction, the polymer beads experience either zero or finite friction
in the pore. In the zero friction case the polymer beads inside the pore are
moved by a constant force included in the molecular dynamics and the
momentum is conserved in the SRD step performed every 50th MD step. In order
to investigate the effect of pore friction the momentum was optionally
destroyed in the SRD step in a manner similar to switching off hydrodynamic
modes in the solvent. This amounts to the polymer beads experiencing finite
friction inside the pore.

\begin{figure*}[htbp]
\centerline{
\includegraphics[angle=0,height=0.17\textheight]{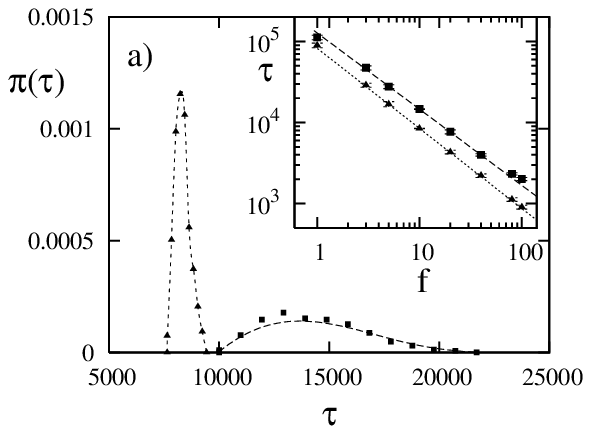}
\includegraphics[angle=0, height=0.17\textheight]{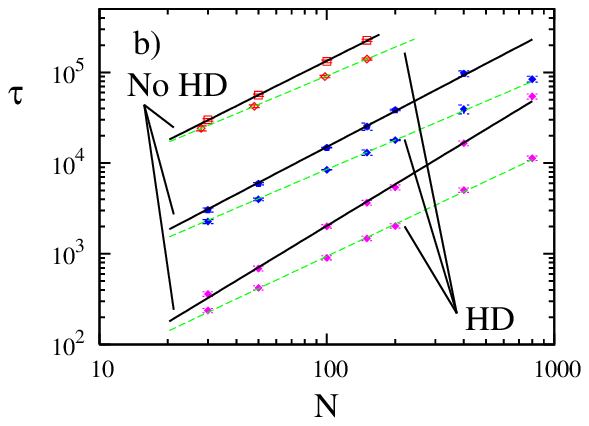}
\includegraphics[angle=0, height=0.17\textheight]{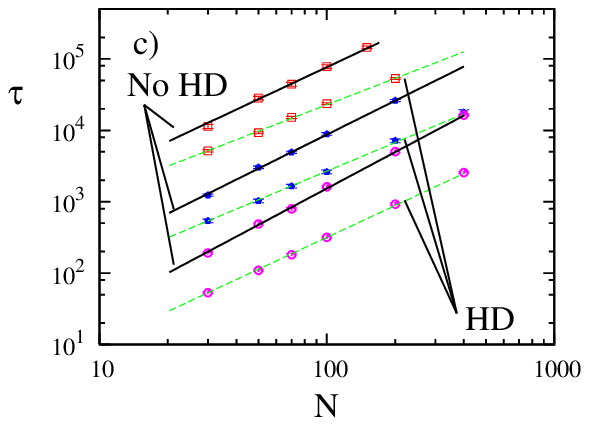}
}
\caption[t_scaling] { 
(Colour online)
a) The distribution of translocation times $\tau$ for
chains of length $N=100$ and a constant pore force $f=10$, with
hydrodynamics ($\blacktriangle$) (averaged over $300$ runs) and without
hydrodynamics ($\blacksquare$) (averaged over $400$
runs). \textbf{Inset}: Average translocation time $\tau$ as a function
of the driving force $f$. The scaling $\tau \sim f^\alpha$ is obtained
with ($\blacksquare$) $\alpha = -0.940 \pm 0.013$ for $f \ge 3$, and
($\blacktriangle$) $\alpha = -0.994 \pm 0.008$ for $f \ge 1$ without
and with hydrodynamics, respectively. The
chain length is $N=100$, and the pore is frictional. b) \& c) Average
translocation time $\tau$ as a function of the chain length $N$ with b)
frictional pore and c) frictionless pore. Results are displayed both with
and without hydrodynamics (HD). b) Shown are $\tau \sim N^\beta$ for forces
$f=1,10,100$ (from top to bottom). For forces $f=1,3,10,40,100$, $\beta =
1.25 \pm 0.02$, $1.26 \pm 0.02$, $1.31 \pm 0.02$, $1.43 \pm 0.03$, $1.52 \pm
0.03$ without HD and $\beta = 1.05 \pm 0.02$, $1.04 \pm 0.01$, $1.07 \pm
0.01$, $1.13 \pm 0.01$, $1.18 \pm 0.02$ with HD.  c) Shown are $\tau \sim
N^\beta$ for forces $f=1,10$ (from top to bottom). For forces
$f=1,3,10,100$, $\beta = 1.50 \pm 0.04$, $1.50 \pm 0.03$, $1.58 \pm
0.03$, $1.70 \pm 0.03$ without HD,
and $\beta = 1.23 \pm 0.03$, $1.26 \pm 0.02$, $1.33 \pm 0.02$, $1.48
\pm 0.02$ with HD. }
\label{t_scaling}
\end{figure*}

Diffusion of the chains and single particles has been verified to satisfy
equipartition theorem in equilibrium. The swelling exponent value of a
self-avoiding-chain, $\nu = 0.6 \pm 0.05$, was measured and the fluctuations
of the radius of gyration the polymer configurations in directions aligned
with and perpendicular to the side walls were measured and found
equal~\cite{linna}. Thus the side walls do not affect the dynamics of
polymers.  Translocation simulations were started from initial
configurations that were checked to be in equilibrium with respect to the
radius of gyration, $R_g$.  The measured $R_g$'s were clearly smaller than
the channel dimensions, which excludes effects due to polymer confinement.
Also, finite (simulation box) size effects were verified not to affect the
dynamics by measuring the relaxation time $\tau_{r} \sim \langle R_g(t)
R_g(0) \rangle$ for polymers of length $N=200$ and various volumes (box
sizes). The relaxation times are of the same order than the largest
translocation times presented in Fig.~\ref{t_scaling}~b). The
simulation box sizes are $[25,32,32]$ for $N \leq 50$, $[32,32,32]$ for $N
\leq 100$, $[40,32,32]$ for $N \leq 200$, $[40,40,40]$ for $N \leq 400$, and
$[60,60,60]$ for $N=800$. The measured radii of gyration are $4.91 \pm 0.01$
for $N=50$, $7.19 \pm 0.01$ for $N=100$, and $10.93 \pm 0.01$ for $N=200$.

\begin{figure}[!ht]
\centerline{
\includegraphics[angle=0, height=0.13\textheight]{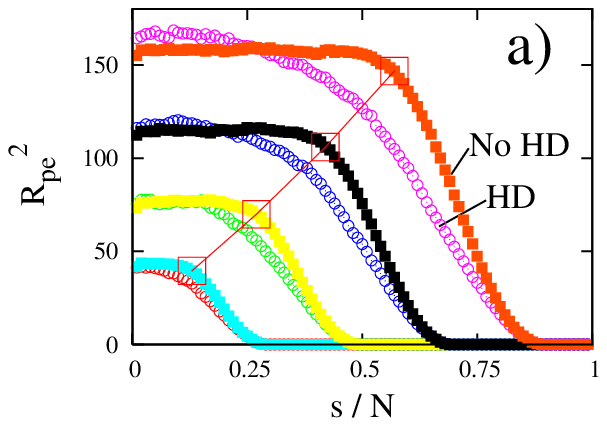}
\includegraphics[angle=0, height=0.13\textheight]{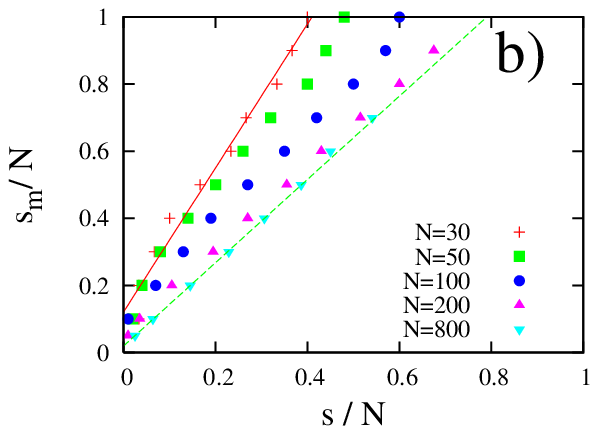}
}
\centerline{
\includegraphics[angle=0,height=0.07\textheight]{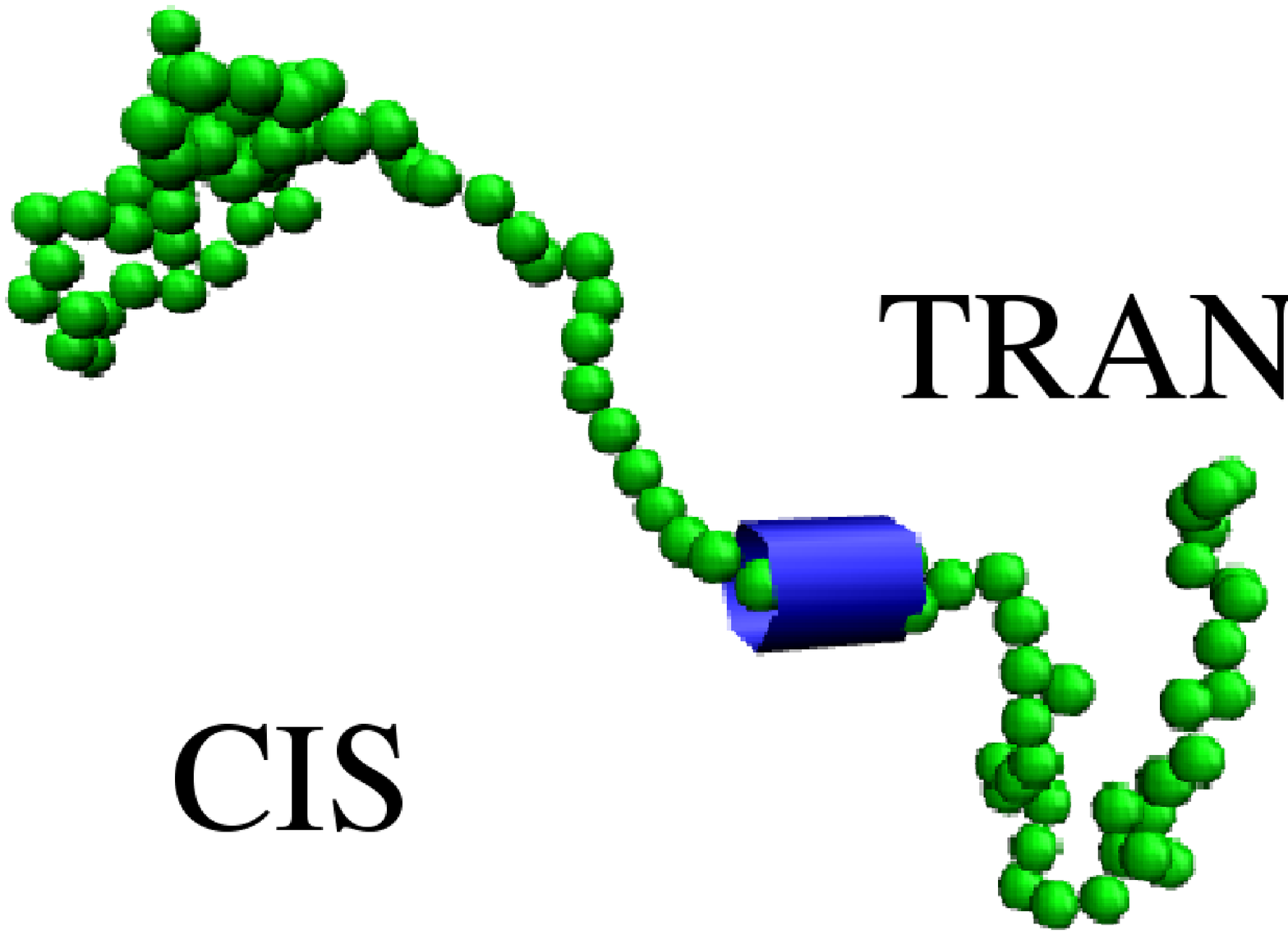}
\includegraphics[angle=0, height=0.14\textheight]{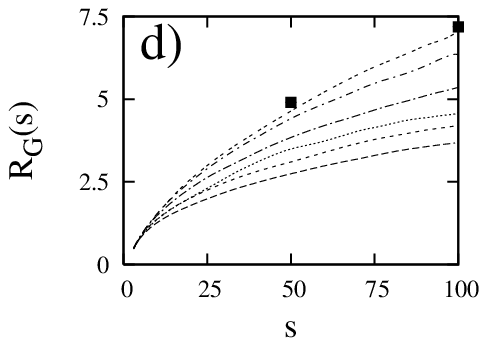}
}
\caption[Rg_trans] {
  (Colour online)
  a) Averaged squared distances of beads numbered $30,50,70,$ and $90$ from the
  pore as a function of the number of translocated beads $s$ for
  polymers of length $N = 100$ with and without hydrodynamics.
  b) The number of mobile beads, $s_m$, (see text) vs number of
  translocated beads,  $s$, both normalised to the polymer length,
  $N$, in the case of no hydrodynamics. $f=10$ in a) and b).
  c) 3D snapshot of a translocating polymer of length $N=100$ at $s=35$. Here
  the pore is frictionless and $f=2$. For clarity reasons, walls are not shown.
  d) The radius of gyration $R_g$ on the {\it trans} side as a function of 
  the number of translocated segments $s$. Chain length $N=100$. The
  applied pore forces for the curves from top to bottom are
  $1,3,10,20,40,$ and $100$. Measured equilibrium $R_g$'s for $s=50$ and
  $s=100$ ($\blacksquare$). The pore is frictional in a), b) and d).
}
\label{Rg_trans}
\end{figure}

In order to characterise the native translocation process, we use pore
force values that are sufficient to induce translocation, hence
obviating any additional constraints for preventing the polymer from
sliding back to the {\it cis} side (see a snapshot of a translocating
polymer in Fig.~\ref{Rg_trans}~c)). As has been noted, additional
constraints can potentially change the observed scaling of the
translocation time $\tau$ with polymer length $N$~\cite{panja}. We compare our
simulated results to experiments~\cite{storm,meller,kasianowicz} and
also to those obtained analytically from the Brownian
translocation framework~\cite{sung,muthukumar} and the numerical
results supporting anomalous diffusion~\cite{kantor,dubbeldam2,gauthier}. The
assumption of the whole or parts (called folds) of the polymer close
to the pore being in equilibrium is crucial to the application of
the aforementioned frameworks. We show that these
assumptions are invalid in
the experimental pore force range and that hydrodynamics has a profound
effect on the forced translocation. This, in turn, we show to be a highly
non-equilibrium process governed by a simple force balance closely
related to the one presented by Storm {\it et al.}, albeit with
important modifications.

We present the results from our simulations in reduced, dimensionless units.
The unit of length is defined as the polymer bond length, $b$, which
corresponds roughly to the Kuhn length of the translocating polymer in
SI-units, $\tilde{b}$. For our freely-jointed chain (FJC) the Kuhn length
can be taken as $\tilde{b} = 2 \lambda_p$, where $\lambda_p$ is the
persistence length, roughly $40\ \text \AA$ for a single-stranded (ss) and
$500\ \text \AA$ for a double-stranded (ds) DNA~\cite{tinland}. In the
simulations the force is exerted on three beads residing simultaneously
inside the pore. The pore force per bead in SI-units, $\tilde{f}$, is
obtained from the dimensionless force per bead, $f$, as $\tilde{f} \equiv f
k_B T/\tilde{b}$. The simulations were performed at $k_BT = 1$, which we
take to correspond to $\tilde{T} = 300\ \text K$. Hence, the
  dimensionless force $f = 1$ corresponds to the total pore force
  $\tilde{f}_{tot} = 3 \tilde{f}$, giving $\tilde{f}_{tot} \approx
  0.12\ \text {pN}$ for ds DNA and $\tilde{f}_{tot} \approx 1.6\ \text
  {pN}$ for ssDNA. A typical
experimentally used potential driving a polymer through the pore for both
the ssDNA in the $\alpha-$HL and dsDNA in the solid state pore is $\sim 120
\ \text {mV}$, which would give a pore force of $\sim 50 \ \text {pN}$ for
ssDNA and $\sim 110 \ \text {pN}$ for dsDNA. When charge reduction due to
Manning condensation is taken into account, the effective force for dsDNA in
the solid-state pore was evaluated to be in the range $20-50 \ \text
{pN}$~\cite{storm,manning}. For ssDNA in an $\alpha$-HL pore the charge
reduction was evaluated to be even more drastic due to confinement in the
pore in addition to the normal charge reduction, giving $\tilde{f} \sim 5 \
\text {pN}$~\cite{meller,sauerbudge}. This would suggest that also the
estimated force for dsDNA could be smaller. The translocation of a polymer
across a pore in a biological membrane involves in addition friction and
{\it e.g.} interaction of the polymer with the pore
proteins~\cite{laan,shariff}, which without detailed information on those
interactions makes exact mapping of the pore force values used in the
simulation to those in experiments impossible. The primary control parameter
is the total pore force, $\tilde{f}_{tot} = M \tilde{f}$, where $M$ is the
number of points on the polymer contour on which the pore force,
$\tilde{f}$, is exerted. On dsDNA these points can be taken to reside at
intervals determined by the nucleotide spacing, which is $3.4\ \text{\AA}$
for dsDNA and $\approx 4\ \text{\AA}$ for ssDNA. The pore force
per bead in the experiments may be estimated as $\tilde{f} = z q^* V/L$,
where the pore potential $V=120 \text{mV}$ and the number of elementary
charges $e$ per nucleotide is $z = 2$ for dsDNA and $z = 1$ for ssDNA. The
effective charge $q^*$ is taken as $e$ for dsDNA~\cite{storm} and $0.1 e$ for
ssDNA due to charge reduction~\cite{sauerbudge}. This gives $\tilde{f}
\approx 1.92 \text{pN}$ for dsDNA and $\tilde{f} \approx 0.37 \text{pN}$ for
ssDNA. Since the length of the solid state pore is $20\ \text{nm}$, $M
\approx 59$, $\tilde{f}_{tot} \gtrapprox 113\ \text{pN}$ for dsDNA, but
could be considerably smaller as pointed out above~\cite{storm}. The length
of the $\alpha$-HL pore is $52\ \text{\AA}$, so $M \approx 13$, giving
$\tilde{f}_{tot} \approx 5\ \text{pN}$ for ssDNA~\cite{meller}. So in
summary, in spite of the intricacies involved in estimating the true force
exerted on the polymer inside the pore, the experimental force magnitudes
are included in the pore force range $f \in [1, 100]$ used in our
simulations and, what is more important with respect to the
observed of out-of-equilibrium effects pertinent to the forced
translocation, the minimum pore force used in our simulations is well under
the minimum pore force magnitudes used in experiments.

For the frictional pore we obtained the translocation probability of $P_{tr}
= 0.12 \pm 0.05$ for $f = 0.25$ that can be taken as a crude estimate for
the minimum force required for generic forced translocation. $P_{tr}$
reached unity at $f \approx 1$. For a frictionless pore $P_{tr} \sim 0.5$
with $f \simeq 1$, and $P_{tr} = 1$ at $f \approx 2$. Experiments on protein
translocation across inner mitochondrial membrane, where the pore is highly
frictional, showed saturation of $P_{tr}$ with the pore potential $80 \
\text {mV}$~\cite{laan}. The potential in related experiments varied from
$150$ to $240\ \text {mV}$~\cite{huang,chen}. Hence, in terms of generic
translocation, the pore force used in our simulations is in a range relevant
for experiments. The simulated pores were non-aqueous. Having solvent inside
the pore may change the minimum pore force value inducing translocation.
This will be investigated in a future publication.  Experimentally, an
average velocity of $\langle v \rangle \sim 1\ \text {cm/s}$ was measured
for a dsDNA translocating across a solid-state pore with a typical potential
of $\sim 120\ \text {mV}$ (corresponding to a pore force of $20-50 \ \text
{pN}$)~\cite{storm,li}. We obtained an average velocity of $v = 0.004$ for
$f=1$, so the dimensionless simulation time unit corresponds roughly to
$\tilde{t} \sim 0.1 \ \mu \text s$. In our translocation simulations SRD
step is performed every 50th MD step, so the solvent and polymer time steps
are $\Delta t_{SRD} = 0.1 \Delta t$ and $\Delta t_{MD} = 0.002 \Delta t$,
respectively. We obtain $\eta \approx 15.776$ for the viscosity of our
model~\cite{kikuchi}. For a polymer of length $N = 100$ we measured radius
of gyration of $R_g \approx 7.198$. Hence we calculate for the Zimm
relaxation time~\cite{doiedwards}, {\it i.e.} the time it takes for the
entire polymer to relax to an entropically and sterically favourable
configuration, $t_z = 0.398\ \eta R_g^3/k_BT \approx 2.4 \cdot 10^3$.
Estimating the friction coefficient for our model~\cite{kikuchi} to be
$\zeta \approx 25$, we obtain for the corresponding (Rouse) relaxation time
without hydrodynamics\cite{doiedwards} $t_r = \zeta (Nb)^2/3\pi^2 k_BT
\approx 8.5 \cdot 10^3$. So, in our model the estimated ratio of relaxation
times with and without hydrodynamics of a polymer of length $N = 100$ is $R
= t_z/t_r \approx 3.5$. We measured $R \approx 1.6$. The single-particle
Reynold's number for this velocity in our model~\cite{kikuchi} is $Re =
0.005$, which is in the relevant regime for physiological solvents.

First we determine the translocation time, $\tau$, as a function of pore
force, $f$. For the scaling exponents $\alpha$ defined as $\tau \sim
f^\alpha$ we obtain $\alpha = {-0.940 \pm 0.013}$ for $f \in
[3,100]$ and $\alpha = {-0.994 \pm 0.008}$ for $f \in [1,100]$
without and with hydrodynamics, respectively, see the inset of
Fig.~\ref{t_scaling}~a). Hence, essentially $\tau \sim f^{-1}$ was
obtained, as was to be expected for force values large compared
with thermal fluctuations. 

The distribution of translocation times, $\pi (\tau)$, for polymers of
length $N=100$ is shown in the main part of Fig.~\ref{t_scaling}~a). Due to
the larger polymer velocities in forced translocation the effect of
hydrodynamics on forced translocation is much more pronounced than what has
been seen with unforced translocation~\cite{ali,gauthier}. We obtain a
reduction in translocation times due to hydrodynamics, which was also seen
by Fyta {\it et al}~\cite{fyta_pre,fyta}. In addition, hydrodynamics not only 
significantly speeds up forced translocation but also reduces the variance
of measured translocation times, which is induced by the long range
correlations due to hydrodynamics, mediating the effect of the pore force
along the polymer contour.

The measured translocation times scale with polymer length, $\tau \propto
N^\beta$, both with and without hydrodynamics, see Figs.~\ref{t_scaling}~b)
and c). It is noteworthy, however, that there exists no single scaling, but
$\beta$ varies with pore force, $f$. In our simulations, $\beta$ starts from
unity and increases with $f$.
At constant $f$, smaller $\beta$ was obtained for the frictional pore. To
distinguish between the change of $\beta$ due to increasing translocation
velocity, $v$, and due to frictional term, when $f$ was increased, scaling
of $\tau$ with $N$ for a pore with no friction was measured and it was found
that the change of $\beta $ was still significant, see Fig.~\ref{t_scaling}.
The experimentally obtained $\beta \approx 1.27$ for a solid state
pore~\cite{storm} would be obtained in our model with a pore force $f
\gtrapprox 3$. Hence, it can be concluded that the change of $\beta$ with
$f$ arises not only from the change in the frictional contribution to the
translocation dynamics, but also because of dynamic changes due to the
change in $v$, which is a clear indication of out-of-equilibrium
effects. In comparison, Fyta {\it et al.}~\cite{fyta_pre} obtained
$\beta = 1.28 \pm 0.01$, and $\beta = 1.36 \pm 0.03$ with and without
hydrodynamics, respectively, for the pore force $f = 1$, which closely
corresponds to $f = 1$ in our simulations as the pore length
in~\cite{fyta_pre} is approximately $3 b$. A pore of very low friction
was used in these lattice Boltzmann (LB) simulations. Accordingly, the
obtained scaling exponent is in fair agreement with ones we have
obtained for the frictionless pore with hydrodynamics. Also the
increase of $\beta$ when hydrodynamics is switched off qualitatively
agrees with our results. Fyta {\it et al.} report the scaling
exponent only for a single pore force magnitude. The verification of
possibly non-universal scaling exponents varying with the pore
force using the lattice Boltzmann method would be most valuable.

Using Langevin dynamics (where hydrodynamics is excluded) we checked
that linear scaling ($\beta = 1$) can be achieved for any constant pore
force with large enough friction. Linear scaling has been
seen with $\alpha$-HL pores, whose diameter is smaller and
friction larger than those of solid state pores. The important notion
is that $\beta$ varies with pore friction even at moderate friction
values, regardless of hydrodynamic interactions.

Most of our results are for a frictional pore, which is the more
realistic. We use pore force magnitudes $f \ge 1$ at which
translocation takes place with very high probability ($P_{tr} \approx
1$) and hence we do not address {\it unforced} translocation. Without
hydrodynamics, the translocation with a frictionless pore approaches
the scaling $\tau \sim N^{1+\nu} \approx N^{1.6}$ as $f$
increases. $\beta = 1+\nu$ is the scaling exponent predicted by the
Brownian translocation framework, independent of $f$. $\beta$ can be
increased further by applying an unrealistically large pore
force. This increase is due to crowding on the $trans$ side, discussed
further below.

When hydrodynamics is allowed, the polymer segments are moved from their
initial equilibrium positions already before actually being pulled by the
tightening polymer contour. This is seen in Fig.~\ref{Rg_trans}~a), where
the squared distance, $R_{pe}^2(n)$, of the polymer bead, labelled $n$,
measured from the pore on the {\it cis} side as a function of the number of
translocated beads, $s$, is shown.  In the absence of hydrodynamics, the
segments towards the free end are seen to remain immobile until they are
pulled towards the pore, whereas due to hydrodynamic interactions the
distance of the labelled bead $n$ from the pore is seen to start decreasing
right from the beginning of the translocation. Hence, the initial
configuration shows less in the translocation, when hydrodynamics is
included. Instead, the configuration on the {\it cis} side continually
changes towards increasingly extended one. Regarding only the dynamics on the
{\it cis} side, for an initially completely extended polymer asymptotically
$\beta \to 1$ as $f$ is increased, which explains the reduction of $\beta$
at constant $f$ when hydrodynamics is applied. Evidently, mechanisms like
squeezing of equilibrated folds through the pore, suggested
in~\cite{dubbeldam,dubbeldam2}, contradict with this mode of motion, where
alignment of segments in the vicinity of the pore is to be expected and,
indeed, seen in the snapshots, see Fig.~\ref{Rg_trans}~c). 

In order to identify the underlying mechanism in the forced translocation,
we extract from the measured distances of the labelled polymer beads from
the pore, $R_{pe}$ the number of mobile beads, $s_m$. We define a labelled
bead as mobile if this measured distance, averaged over several runs,
changes appreciably. The number of mobile beads, $s_m$, as a function of
translocated beads $s$ is read off from the inflection points in
Fig.~\ref{Rg_trans}~a) depicting the measured $R_{pe}^2(s/N)$ in the case of
no hydrodynamic interactions. In Fig.~\ref{Rg_trans}~b) $s_m$ is plotted as
a function of translocated beads $s$ when hydrodynamics is not included.
Linear dependence $s_m = k s$ is obtained. Up to lengths of $N \approx 200$,
$k \sim N^{-\chi}$ and levels off to a constant value $> 1$ for longer
polymers. At all times, the drag force, $f_d$, balances with the constant
pore force. $f_d$ is exerted on mobile beads, so in the absence of
hydrodynamics $f_d \sim s_m \langle v \rangle$, where $\langle v \rangle$ is
the average velocity of the mobile beads. When the whole chain has
translocated, $f_d \sim N_m \langle v \rangle$, where $N_m = k N$. With no
hydrodynamics, the beads are set in motion from their equilibrium positions,
so the distance $d$ of the mobile bead furthest from the pore scales as $d
\sim N^\nu$. The average translocation time then scales as $\tau \sim
\langle d \rangle / \langle v \rangle \sim k N^{1+\nu} \sim N^{1+\nu-\chi}$.
For the data in Fig.~\ref{Rg_trans}~b), where $f = 3$, we obtain $\chi
\approx 0.35$ that accords with the measured $\beta = 1.26$, see 
Fig.~\ref{t_scaling}~b). With the pore force $f = 100$ the $k$'s for $s_m =
ks$ are smaller and the measured $s_m$-$s$ curves for different $N$ appear
more aligned. Asymptotically, $k \to 1$, $\forall N$, as $f \to \infty$,
{\it i.e.} polymer beads are translocated at the same rate that they are set
in motion. Removing the friction from the pore also makes $k$ values smaller
and more identical for different $N$ due to translocation becoming faster.
Both the increase in the pore force and reduction in the pore friction take
the scaling exponent $\beta$ toward $1+\nu$ due to $s_m$ and hence the drag
force, $f_d$, remaining more constant throughout the translocation.
Hydrodynamics changes the form of the drag force. $f_d$ no more depends
strictly linearly on $s_m$ for configurations of moving polymer segments,
but all beads are set in motion in the beginning of translocation. Yet, the
above described mechanism is still clearly underlying the translocation also
when hydrodynamics is included, see Fig.~\ref{Rg_trans}~a). For a constant
$f$ hydrodynamics reduces $\beta$ due to enhancing collective motion of the
polymer towards the pore. Hence, the initial equilibrium positions do not
determine $\beta$ like in the absence of hydrodynamics.

In addition to the above described dynamics on the {\it cis} side, there is
a potential contribution from the crowding of the polymer beads close to the
pore on the {\it trans} side. This indeed can be seen from the snapshot in
Fig.~\ref{Rg_trans}~c) and the measured radii of gyration, $R_g$, of the
translocated parts of the polymer on the {\it trans} side, which are clearly
smaller than the corresponding equilibrium $R_g$ even for the smallest force
values, see Fig.~\ref{Rg_trans}~d). The crowding is more enhanced for longer
polymers thus increasing $\beta$, in agreement with our measured
translocation times, Figs.~\ref{t_scaling}~b) and c). Asymptotically, for $N
\to \infty$, the average velocity of the polymer beads on the {\it cis}
side, $\langle v \rangle$, would have to diminish as $s_m$ increases to
maintain the force balance $f_d = f_p = \text {constant}$. Eventually $s_m$
would be so large that the polymer barely moves, which would bring the
translocation to the regime where diffusive motion of the beads shows. For
finite polymers this is not the case. Instead, the simulated polymers whose
lengths measured in Kuhn lengths are well in the range of polymers used in
experiments clearly show that the moving polymer segment does not slow down
to velocities where diffusive motion could be seen. Crowding was also
reported in the LB simulations in~\cite{fyta_pre}. However, the alignment of
segments in the vicinity of the pore, Fig.~\ref{Rg_trans}~c), was not
observed, but on the contrary polymers were reported to stick to the wall on
the {\it cis} side.

In conclusion, we have studied forced polymer translocation by a model where
hydrodynamics is taken judiciously into account. In our minimal
model the pore is non-aqueous, which precludes hydrodynamic coupling
of the two chambers separated by the wall, and the pore potential is
the only driving
force for the translocation. No additional mechanisms for preventing the
translocated polymer segments from sliding back to the {\it cis} side were
included. Using this model, the smallest pore force at which
polymers translocate was estimated to be in the order of $1 \ \text{pN}$,
according with experimental findings~\cite{storm,li,laan}. The used pore
force values cover the biologically relevant range and thus characterise
well the essential dynamics of forced translocation in biological systems,
and DNA experiments. Hydrodynamics was shown to significantly speed up the
translocation and diminish variation in the translocation times. The scaling
exponent, $\beta$, of the translocation time with respect to polymer length
was seen to increase with pore force, $f$. The obtained scaling exponents
and their variation with the pore force could be explained by simple force
balance at the pore and the observation that the rate, at which the size of
the part of the polymer in {\it cis} side moving toward the pore grows with
respect to the part translocated to the {\it trans} side, varies with $f$. A
simple estimate was given for the case when hydrodynamics is not included,
which agreed with the numerical results.  Consequently, a single universal
exponent cannot describe translocation for all $f$. The magnitude of pore
friction was also seen to change $\beta$. Linear scaling, $\beta =1$, was
obtained for large enough friction. $\beta$ was shown to change also due to
crowding mechanism on the {\it trans} side. As crowding increases with pore
force it is an additional mechanism accounting for the increase of $\beta$
with $f$. In summary, by using realistic dynamics where hydrodynamics is
included we have shown that experimentally observed forced translocation can
be described by a simple force balance. The forced translocation process was
also shown to be a highly non-equilibrium process for the experimentally
relevant force regime, which explains the discrepancy between theoretical
approaches and experiments. We have shown that no universal scaling of the
translocation time with the polymer length exists. Experimental verification
of this by using different pore potential magnitudes would be very
important.

\acknowledgments{
This work has been supported by the Academy of Finland (Project No.~127766).
}

\end{document}